# Unraveling the Electronic Structure of Narrow Atomically-Precise Chiral Graphene Nanoribbons

*Néstor Merino-Díez [‡, †], Jingcheng Li [†, §], Aran Garcia-Lekue [‡, ⊥], Guillaume Vasseur [‡, §], Manuel Vilas-Varela[∇], Eduard Carbonell-Sanromà [†], Martina Corso [†, §], J. Enrique Ortega [‡, §, ⁂], Diego Peña[∇], Jose I. Pascual [†, ⊥], and Dimas G. de Oteyza\* [‡, †, §, ⊥].*

[‡]Donostia International Physics Center (DIPC), 20018 San Sebastián-Donostia, Spain

[†]CIC nanoGUNE, 20018 San Sebastián-Donostia, Spain

[§]Centro de Física de Materiales (CSIC/UPV-EHU) - PPC, 20018 San Sebastián-Donostia, Spain

[⊥]Ikerbasque, Basque Foundation for Science, 48013 Bilbao, Spain

[∇]Centro Singular de Investigación en Química Biolóxica e Materiais Moleculares (CIQUS) and Departamento de Química Orgánica, Universidad de Santiago de Compostela, 15782 Santiago de Compostela, Spain

[⁂]Departamento de Física Aplicada I, Universidad del País Vasco (UPV/EHU), 20018 San Sebastián-Donostia, Spain

AUTHOR INFORMATION

**Corresponding Author:**

*E-mail: d_g_oteyza@ehu.es

ABSTRACT

Recent advances in graphene nanoribbon-based research have demonstrated the controlled synthesis of chiral graphene nanoribbons (chGNR) with atomic precision using strategies of on-surface chemistry. However their electronic characterization, including typical figures of merit like band gap or frontier band´s effective masses, has not yet been reported. In this work, we provide a detailed characterization of (3,1)-chGNRs on Au(111). The structure and epitaxy, as well as the electronic band structure of the ribbons, are analyzed by means of scanning tunneling microscopy and spectroscopy, angle resolved photoemission and density functional theory.

**TOC GRAPHICS**

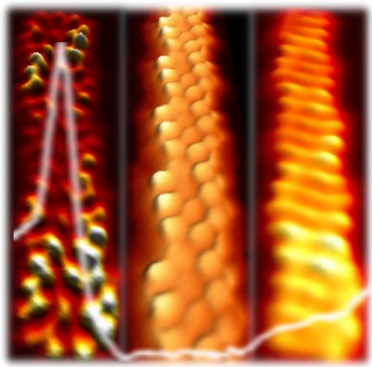

The growth and characterization of new atomically precise graphene nanoribbon (GNR) structures is a challenging quest. The research efforts toward that goal are continuously increasing, driven by the promising prospects of GNR-based technologies.[1,2] As a result, a relatively large variety of armchair graphene nanoribbons (aGNR) has already been synthesized on different coinage metal surfaces.[3–7] Their subsequent characterization has proved the predicted band gap dependence on the ribbon width to be true.[8–10] Also zigzag GNRs (zGNR) have been successfully synthesized from adequate molecular precursors,[11] further proving the presence of the highly coveted edge states associated to zigzag edges.[11–13] However, graphene nanoribbons with chiral edge orientations, i.e. with periodically alternating armchair and zigzag segments, have been hardly characterized to date.[14,15] The first report on the synthesis of an atomically precise chiral GNR (chGNR) came from a surprising result in which a precursor designed to render aGNR resulted in chGNRs when deposited on a Cu(111) surface.[16] This unexpected reaction path arises from a very specific molecule-substrate interaction and was studied in detail later on.[17,18] However, although some spectroscopic measurements were performed on such Cu(111)-supported ribbons,[19] important figures of merit like the band gap value or the frontier band´s effective masses remain unknown.

Recently we reported the design of an alternative precursor molecule (2,2'-dibromo-9,9'-bianthracene) that resulted in the formation of (3,1)-chGNRs independently of the substrate used, at least on the explored Au(111), Ag(111) and Cu(111) surfaces.[20] Thus, in addition to the advantages in the growth process that lead to longer chGNRs at reduced processing temperatures, it places at our disposal chGNRs on a weaker interacting surface like Au(111). On such a surface, hybridization effects with the substrate are weaker and the ribbon´s properties easier to probe. In this work we have made use of this advantage, studying the structural and electronic properties of (3,1)-chGNRs on Au(111) and Au(322) by scanning tunneling microscopy and spectroscopy (STM and STS), angle resolved photoemission spectroscopy (ARPES), and density functional theory (DFT).

The precursor molecule and the two-step reaction path toward the final chGNR are displayed in Fig. 1a.[20] In a first step, Ullmann coupling of the surface-supported precursors sets in at temperatures above 140ºC, leading to non-planar polymeric structures due to the steric hindrance exerted mainly by hydrogen atoms placed within the anthracene units. In a second step, cyclodehydrogenation of the polymeric structures ends up in planar chGNRs (Fig 1d,e), formed entirely by sp$^2$ carbon atoms, saturated with single H atoms along the edges. As reported earlier,[20] the strained structure of the polymer lowers the cyclodehydrogenation threshold temperature below 200 ºC, allowing to obtain these ribbons at temperatures much lower than most other GNRs published to date.[3,4,7,11]

Characterization of the GNR structure and distribution has been performed by STM. Low-temperature STM (LT-STM) using CO-functionalized tips allows achieving high intramolecular resolution when scanning at short tip-sample distances within the Pauli repulsion regime.[21–24] As displayed in Fig. 1e, we have made use of this effect to resolve the nanoribbon´s internal bonding structure. From larger scale images, it becomes evident that the ribbons display six well-defined preferential orientations (Fig. 1b), each with the ribbon´s axis deviated ~16º from the [10-1] (and equivalent) substrate directions. The distribution of chGNR orientations is plotted in the inset of Fig. 1b, including the high-symmetry substrate directions as a reference. Given the chiral nature of the ribbons, the six orientations correspond to three substrate-related azimuthally equivalent directions for each of the two enantiomeric structures (marked by blue and red in Figure 1, respectively). The associated epitaxial model extracted from high resolution images is displayed in Fig. 1c, showing commensuration at every second unit cell of the chiral ribbons.

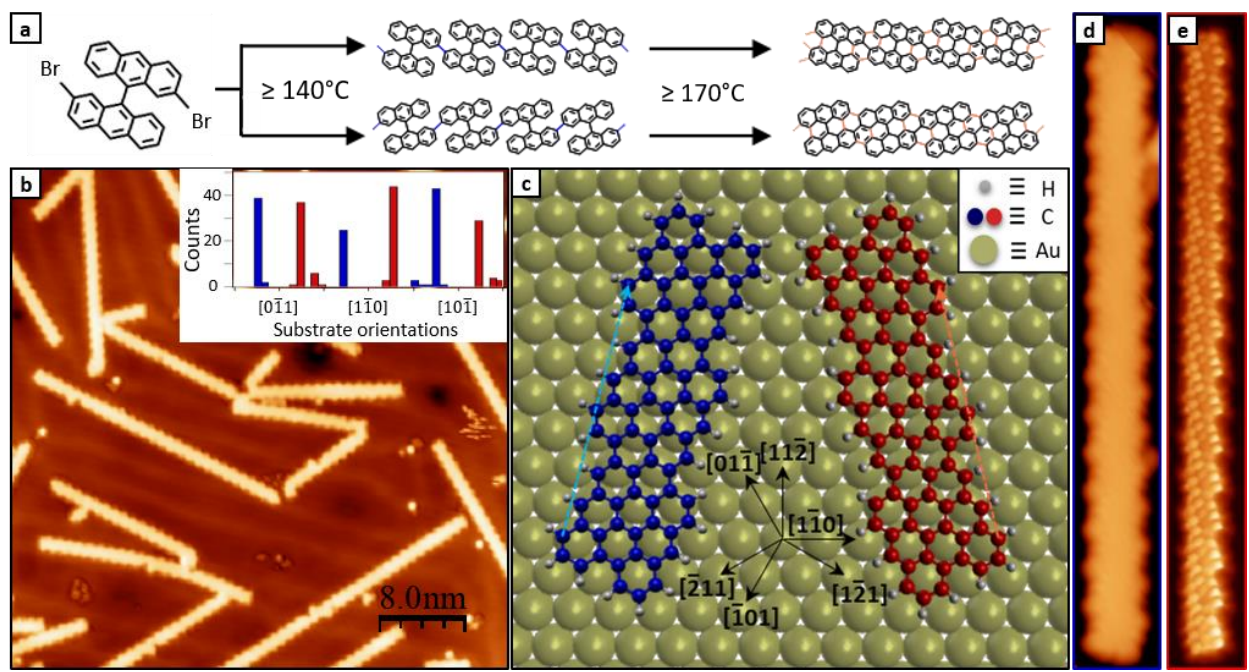

**Figure 1.** Synthesis, structure and epitaxy of (3,1)-chGNRs on Au (111). (a) Schematic reaction path for the synthesis of (3,1)-chGNR with threshold temperatures indicated for each synthetic step. (b) Constant current STM image (45 nm x 45 nm; $V_s$ = -0.15 V; $I_t$ = 0.05 nA) of a representative (3,1)-chGNRs sample on Au(111) after annealing to 350 ºC, with the histogram on the azimuthal orientation distribution with respect to the high-symmetry substrate directions (inset) obtained from the analysis of 245 different nanoribbons. (c) Epitaxial relation exemplified with three-monomer-long (3,1)-chGNRs enantiomers on Au(111), where blue/orange dashed arrows depict the commensuration every two unit cells. The translational adsorption site in the model is arbitrary, since the particular adsorption position could not be unambiguously extracted from the experimental images. (d) Constant current STM (15.4 nm x 2.5 nm; $V_s$ = -1.1 V; $I_t$ = 0.11 nA) and (e) contant height STM image (15.4 nm x 2.5 nm; $V_s$ = 2 mV) obtained with a CO-functionalized tip. Red/blue colors are employed to specify the information associated with each enantiomer in figure b (inlet), c, d and e.

The electronic properties of the ribbons have been first characterized by STS. Fig. 2a displays a conductance point spectrum on a ribbon, together with a reference spectrum on the surrounding substrate. Because tunneling conductance is proportional to the local density of states (LDOS) at the probe position, one can clearly distinguish the onset of the ribbon´s valence band (VB) and conduction band (CB). From a statistical analysis of several tens of ribbons we find the band´s onsets at -0.22 ± 0.05 V and 0.45 ± 0.02 V, respectively. The resulting band gap of 0.67 ± 0.06 eV is larger than that obtained from DFT calculations (Fig. S1, note that the undersestimation of band gaps is a well-known limitation of DFT). However, the constant-height conductance maps at the onset energies (Fig. 2b,c) show excellent agreement with the calculated wave functions (Fig. 2d,e) of the frontier states of valence and conduction band at the gamma point (in spite of being measured with a CO-functionalized tip[25]), providing further confirmation on the nature of those experimentally measured states. This is also confirmed by constant-current conductance maps with a non-functionalized metallic tip (Fig. S2), where clear GNR-related density of states appears as the energy reaches either band onset, evidencing similar patterns as those in Fig. 2. Those patterns are clearly different for VB and CB, the latter appearing with a characteristic wave-front structure, while the former displays a more complex sequence of lobes (Fig. 2).

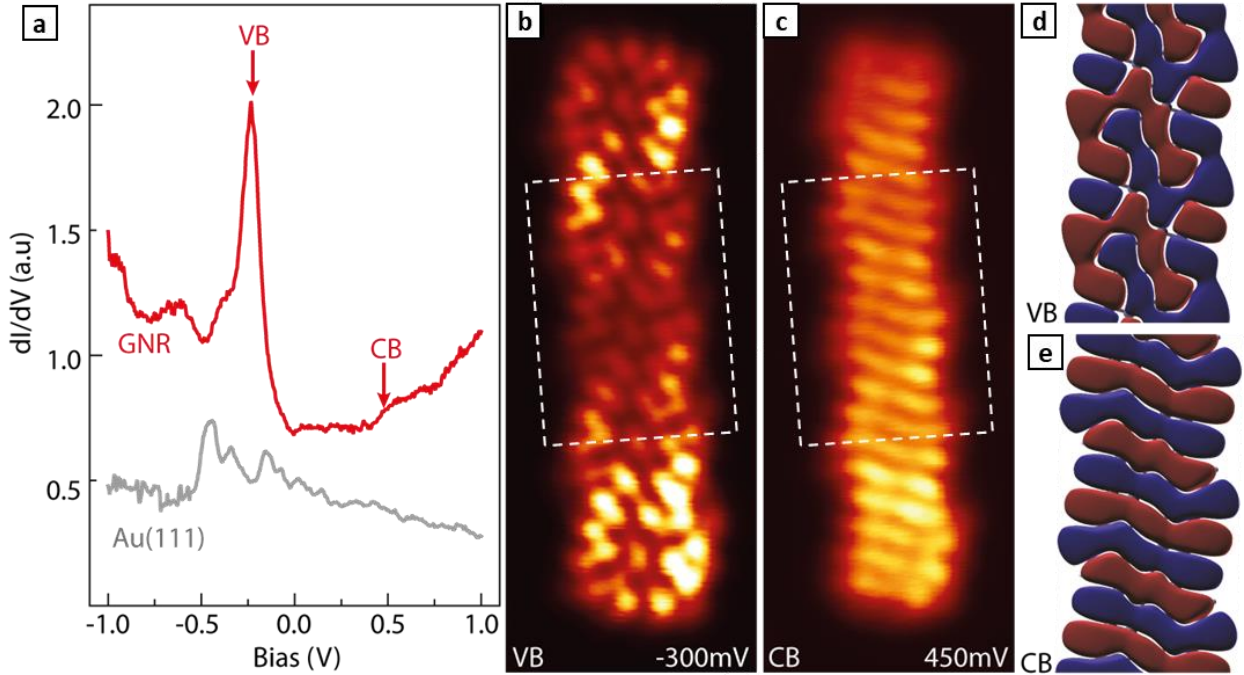

**Figure 2.** Spectroscopic characterization of frontier molecular orbitals of (3,1)-chGNRs on Au(111). (a) Representative dI/dV point spectra obtained from (3,1)-chGNRs on Au(111) (in red) with Au(111) signal (in grey) included as background reference (open-feedback parameters: $V_s$ = 1.0 V, $I_t$ = 0.5 nA, modulation voltage $V_{rms}$ = 0.1 mV). (b,c) STM constant-height conductance maps (2.0 nm x 5.2 nm; open-feedback parameters: $V_s$= 0.2V; $I_t$ = 0.06nA; $V_{rms}$= 1mV) near the (b) valence (-300mV) and (c) conduction (450 mV) band onsets. (d,e) DFT simulations of the wave function for states at the onset of (d) valence and (e) conduction bands (at the Γ point), on an area equivalent to the dashed rectangle in (b) and (c) respectively. Red and blue colors represent isosurfaces of positive and negative wave function amplitudes, for an isovalue of 0.015 Å$^{-3/2}$.

Both conduction and valence bands display a dispersive behavior as they deviate from Γ (Fig. S1). A fingerprint of it is found in conductance maps over a wider energy range of the VB, revealing an additional energy-dependent LDOS modulation along the ribbon axis (Fig. 3a-f). To quantify this effect, we measured equidistant point spectra along the edge of a GNR (Fig. 3g), displayed in Fig. 3h as a function of its position along the ribbon with a color coded conductance intensity (z-axis). In addition to the edge periodicity arising from the chGNR structure, another energy-dependent modulation appears. It coincides with that observed in the conductance maps, in which the number of nodes increases as the energy departs from the band onset. It relates to the formation of standing waves from electronic states scattered at the nanoribbon edges, thus holding the band´s dispersion relation information. This can be distinguished best in Fig. 3i, which depicts a line-by-line Fourier transform (FT) of Fig. 3h, and thus the dispersion of the probed bands.

The VB is observed dispersing down with an effective mass of -0.34 ± 0.05 $m_0$, as obtained from a parabolic fit to the topmost region of the band. In contrast, no dispersion information has been obtained for the CB from the FT-STS analysis. Indeed, the CB is much harder to detect in STS measurements, as can already be guessed from the marked asymmetry in the signal strength of the STS spectrum in Fig. 2a for VB and CB. As explained in detail in previous works,[7,26] the faster a wave function changes its sign along the ribbon axis, the lesser it extends into the vacuum along the GNR normal. This makes it less accessible to STM/STS measurements, where tip-sample distances remain typically above 5 Å. Our wave function calculations of (3,1)-cGNRs in Fig. 2d,e reveal the CB to change sign along the ribbon axis faster than the VB, thus agreeing with its poorer detection in our spectral measurements.

Interestingly, additional features in the FT-spectral map of the VB are observed, not present in previous works performing a similar analysis on aGNRs.[7,26] A replica of the dispersive band appears shifted by 3.5 nm$^{-1}$ (displayed with a green dashed line in Fig. 3i), namely centered at the Brillouin zone edge, which is defined by the GNR periodicity arising from its chiral edges [periodicity $a$ = 8.97 Å = $\pi$ / (3.5 nm$^{-1}$)]. This band replica and the increased intensity line at the zone edge can be traced back to the additional modulation from the GNR chirality. The imposed periodicity stresses the Bloch wave function character of the electronic states, whose coherent addition resulting from scattering events has been previously shown to lead to exactly those two types of features in FT-STS.[27] The periodicity of these features in reciprocal space can be additionally observed in the line-by-line FT-spectra plotted over a wider energy and momentum range displayed in Fig. S3.

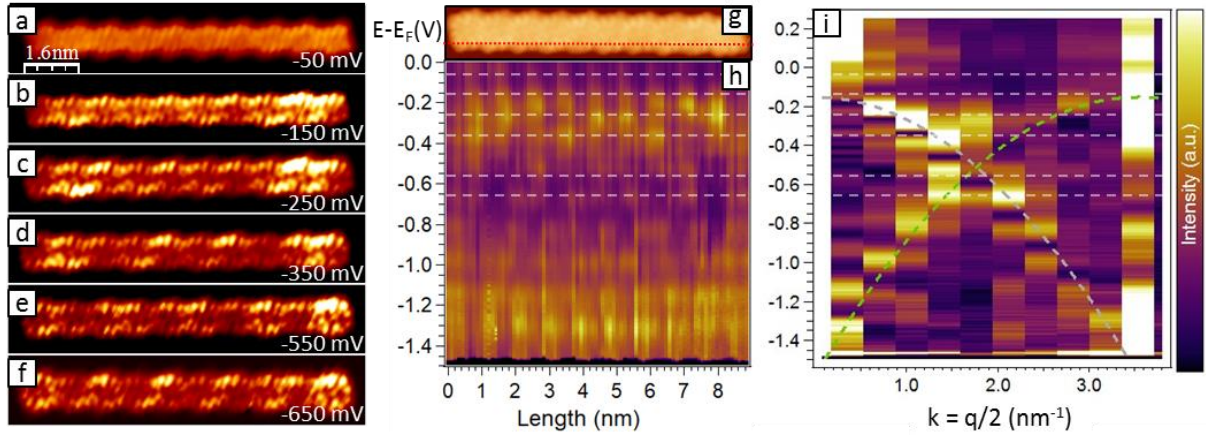

**Figure 3.** STM constant-height conductance maps (10.3 nm x 2.0 nm; $I_t$ = 0.03 nA; modulation voltage $V_{rms}$ = 12mV ) (a) within the band gap at -50mv, (b) near the valence band onset at -150mV, (c) at -250mV, (d) at -350 mV, (e) at -550 mV and (f) at -650 mV, exhibiting confined standing waves along the ribbon. (g) Constant current STM image of the same ribbon, showing the path (red dashed line) followed by the equidistant dI/dV spectra. (h) Color-coded conductance signal obtained from equidistant dI/dV point spectra (open-feedback parameters: $V_s$ = 1.50 V; $I_t$ = 0.8nA; $V_{rms}$= 12 mV) on the ribbon and along the red dashed line displayed in (g). (i) Line-by-line Fourier transform from the stacked spectra in (h), showing the two-parameter parabolic fit (grey dashed line) used for extracting the effective mass. The additional parabola centered around the Brillouin zone edge is displayed with a green dashed line. Grey horizontal lines corresponding to the voltage biases of maps in (a-f) are superimposed in (h) and (i) as a guide to the eye.

In order to compare the VB dispersion properties obtained from FT-STS with results from a more standard approach we have characterized the (3,1)-cGNRs also by angle resolved photoemission spectroscopy (ARPES). Similar comparisons have been performed previously on the VB dispersion of 7-aGNRs and 9-aGNRs. In the former, the effective mass extracted from FT-STS and ARPES differed by a factor 2,[26,28] while in the latter both techniques were in agreement within error bars.[7] Since ARPES is an ensemble averaging technique, having uniaxially aligned GNRs is a requirement to measure the dispersion along a well-defined direction. The aligned growth of aGNRs has been successfully achieved by using a Au(788) surface as template,[28,29] which features ~4 nm wide (111) terraces periodically separated by steps running along the compact [10-1] direction. However, because the chiral GNRs studied here display a markedly preferred growth orientation at 16º off from the compact [10-1] (and equivalent) direction (see Fig. 1), the Au(788) terraces do not satisfactorily guide an uniaxial growth of the ribbons. Instead, the growth results in low quality samples with short ribbons oriented partially along the step edges, but also along their epitaxially favored directions (Fig. S4). The scenario changes when using narrower terraces. The ~1.2 nm wide terraces of Au(322) are wide enough to host a (3,1)-chGNR, but narrow enough to largely inhibit molecular coupling along any other orientation than following the terraces. As a result, uniaxially aligned ribbons could be grown on Au(322), as shown in Fig. 4a-b.

The subsequent ARPES characterization is displayed in Fig. 4c (associated raw data are shown in Fig. S5). While no GNR signal is observed in the first Brillouin zone and only a weak shadow in the second, the VB is nicely resolved in the third Brillouin zone. From a parabolic fit to the topmost VB region we extract values of -0.50 ± 0.02 eV and -0.36 ± 0.04 $m_o$ for the band´s onset energy and effective mass, respectively. Comparing to the results from FT-STS, the effective mass shows agreement within error margins (-0.36 ± 0.04 $m_0$ from ARPES vs. -0.34 ± 0.05 $m_0$ from FT-STS), but the band onset is notably lower in energy (i.e. -0.5 ± 0.02 eV from ARPES vs. -0.22 ± 0.05 mV from STS). This, however, can be easily explained by the different nature of the probed samples. The large step density of Au(322) lowers its work function with respect to that of Au(111). As extracted from measurements of the cut off energy of photoemitted electrons from either surface (Fig. S6) the work function changes by 0.25 eV, fitting well with the measured difference in VB onsets (~0.28 eV). A vacuum level change rigidly shifting down the adsorbate´s band structure in a simple vacuum level pinning scenario readily explains the offset and provides a fully coherent scenario for the comparison of STS and ARPES data.

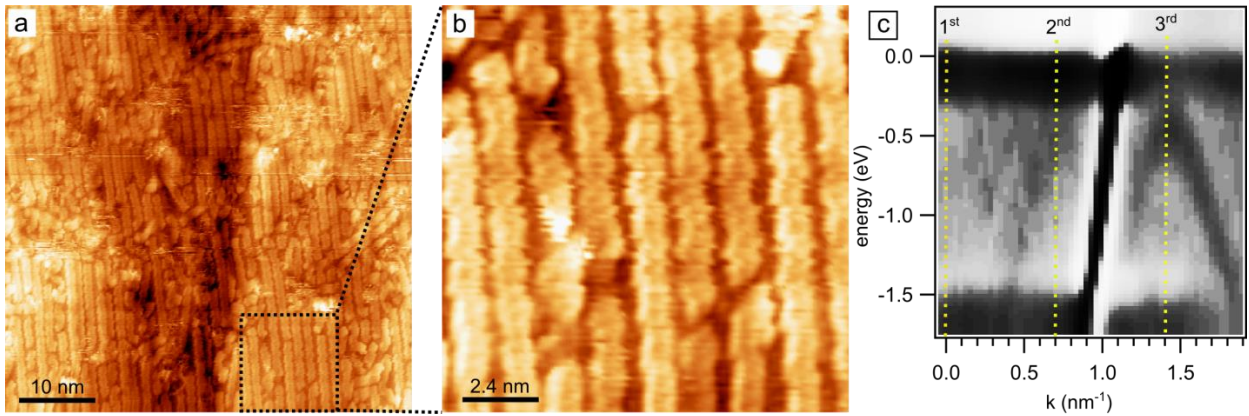

**Figure 4.** (a) Large scale constant current STM topography image (50 nm x 50 nm; $V_s$ = -0.2V; $I_t$ = 0.02 nA) and (b) zoom in (10 nm x 10 nm) for an easier appreciation of details, of (3,1)-chGNRs on Au(322). (c) Second derivative of the photoemission intensity reflecting the valence band dispersion (raw photoemission intensity data are displayed in the supplementary information). Dotted lines mark the center of first, second and third Brillouin zones.

At this point it is interesting to compare the electronic properties of (3,1)-chGNRs and of 7-aGNRs. Both ribbons have comparable widths and result from precursors sharing the same carbon backbone, although polymerizing along different directions. However, the change in edge orientation from armchair to a chiral (3,1) direction brings about dramatic changes in the electronic properties. By way of example, the band gap is reduced from 2.37 ± 0.06 eV in the former [26] to 0.67 ± 0.06 eV in the latter, although without evident signatures of the spin polarized edge states predicted to appear around the Fermi level in chiral ribbons.[30–33] This is presumably due to a too small GNR width and the associated band gap opening.[30-32] According to calculations, a closing of the bandgap and the appearance of edge states are predicted to occur either increasing the GNR width or also maintaining a similar GNR width but with chiralities closer to the zigzag direction (as well as for pure zGNRs).[30-32] However, an experimental confirmation of such behavior is still missing.

In the intuitive picture of GNR bands being directly related to the dispersion properties of the parent material graphene (as happens with aGNRs),[12] the effective mass is expected to correlate with the band gap.[34] Indeed, experimental data revealed that a band gap drop from 2.4 eV to 1.4 eV going from 7-aGNRs to 9-aGNRs brings about a substantial effective mass reduction from 0.21 $m_0$ (obtained from ARPES)[28] or 0.41 ± 0.08 $m_0$ (obtained from FT-STS)[26] to ~0.1 $m_0$ (obtained from FT-STS and ARPES alike).[7] However, that scenario gets much more complicated as GNRs with different edge orientations are compared, whereby the GNR bands are no longer trivially related those of graphene.[12] As a result, our DFT calculations predict only a minor decrease of the effective mass going from 7-aGNRs (0.33 $m_0$ [35]) to (3,1)-chGNRs (0.27 $m_0$) in spite of the greatly diminished band gap from 2.37 eV to 0.67 eV. Experimentally the effective mass changes from 0.21 $m_0$ (ARPES)[28] or 0.41 ± 0.08 $m_0$ (FT-STS) in 7-aGNRs[26] to ~0.35 $m_0$ (FT-STS and ARPES) in (3,1)-chGNRs, confirming the dramatically different behavior of GNRs with different edge orientations.

In conclusion, we have provided a thorough characterization of the structural and electronic properties of (3,1)-cGNRs on Au(111). A strong favoritism for commensurate adsorption directions is observed that can, however, be overcome with the use of adequately stepped surfaces that prevent the ribbon growth along any other direction than following the terraces. Such samples have been used to characterize the band dispersion by ARPES and to compare the results with those obtained from Fourier transform scanning tunneling spectroscopy measurements. We end up with a fully coherent picture of the GNR´s band gap ($0.67 \pm 0.06$ eV), effective mass (~$0.35\,m_0$) and energy level alignment (shifting with the substrate work function as in an ideal vacuum level pinning scenario) that will enable a better understanding of their performance in future electronic devices and allow a rational design of heterostructures with complementary GNRs.

METHODS

For the preparation of the different samples, 2,2'-dibromo-9,9'-bianthracene molecular precursor was sublimated at ~425 K from a Knudsen cell and oriented to metallic substrates for deposition. Atomically cleaned surfaces Au(111) and Au(322) were achieved by standard sputtering and annealing cycles. Measurements on Au(111) were performed on a home-build, low-temperature STM under ultra-high vacuum (UHV), that is at pressures below $10^{-10}$ mbar and a base temperature of 4.8 K. Measurements on Au(322) where performed in a UHV system combining a commercial Omicron VT-STM connected to a homebuilt ARPES system equipped with a closed-circuit He-compressor-cooled manipulator, a monochromatized gas discharge lamp and a SPECS Phoibos 150 electron analyzer. STM and ARPES measurements could thus be performed sequentially on the same sample without breaking the UHV conditions. ARPES measurements were performed at a sample temperature of 90 K using the He I line (21.2 eV). All STM images were processed by WSxM software.[36]

The structural and electronic properties of free-standing (3,1)-cGNRs were calculated using density functional theory (DFT), as implemented in the SIESTA code.[37] We considered a supercell consisting of a (3,1)-chGNR infinite along the x axis, with vacuum gaps of ~15 Å in y and z directions in order to avoid interactions between ribbons in adjacent cells. A Monkhorst-Pack k-point grid with 101x1x1 k-points was used for the Brillouin zone sampling and the mesh cut-off for real space integrations was set to 300 Ry. A basis set consisting of split-valence double-zeta plus polarization DZP orbitals was employed, and a variable-cell relaxation of the periodic system was performed until residual forces on all atoms were less than 0.01 eV/Å. Dispersion interactions were taken into account by the non-local optB88-vdW functional.[38]

ASSOCIATED CONTENT

Constant current conductance maps evidencing the absence of signal at energies within the gap and the clear GNR-related signal at the VB and CB onsets. DFT calculated band structure for (3,1)-chGNRs. Line-by-line FT-STS data displayed in Fig. 3i depicting the complete energy and momentum range experimentally measured. Comparative STM images of (3,1)-chGNR growth on Au(788) and Au(322). ARPES measurements displaying the raw photoemission data of the GNR-related dispersive band. Work function measurements of Au(111) and Au(322) surfaces.

AUTHOR INFORMATION

**Corresponding Author:**

*E-mail: d_g_oteyza@ehu.es

**The authors declare no competing financial interests.**

ACKNOWLEDGMENT

The project leading to this publication has received funding from the European Research Council (ERC) under the European Union's Horizon 2020 research and innovation programme (grant agreement No. 635919), from the Spanish Ministry of Economy, Industry and Competitiveness (MINECO, Grant Nos. MAT2016-78293-C6, FIS2015-62538-ERC), from the Basque Government (Grant Nos. IT-621-13, PI-2016-1-0027, PI-2015-1-42, PI-2016-1-0027); from the European Commission in FP7 FET-ICT "Planar Atomic and Molecular Scale devices" (PAMS) project (Contract No. 610446), from the Xunta de Galicia (Centro singular de investigación de Galicia accreditation 2016−2019, ED431G/09), and from the European Regional Development Fund (ERDF).

REFERENCES

(1) Bonaccorso, F.; Colombo, L.; Yu, G.; Stoller, M.; Tozzini, V.; Ferrari, A. C.; Ruoff, R. S.; Pellegrini, V. Graphene, Related Two-Dimensional Crystals, and Hybrid Systems for Energy Conversion and Storage. *Science* **2015**, *347*, 1246501–1246501.
(2) Celis, A.; Nair, M. N.; Taleb-Ibrahimi, A.; Conrad, E. H.; Berger, C.; Heer, W. A. de; Tejeda, A. Graphene Nanoribbons: Fabrication, Properties and Devices. *J. Phys. Appl. Phys.* **2016**, *49*, 143001.
(3) Cai, J.; Ruffieux, P.; Jaafar, R.; Bieri, M.; Braun, T.; Blankenburg, S.; Muoth, M.; Seitsonen, A. P.; Saleh, M.; Feng, X.; *et al.* Atomically Precise Bottom-up Fabrication of Graphene Nanoribbons. *Nature* **2010**, *466*, 470–473.
(4) Chen, Y.-C.; de Oteyza, D. G.; Pedramrazi, Z.; Chen, C.; Fischer, F. R.; Crommie, M. F. Tuning the Band Gap of Graphene Nanoribbons Synthesized from Molecular Precursors. *ACS Nano* **2013**, *7*, 6123–6128.
(5) Abdurakhmanova, N.; Amsharov, N.; Stepanow, S.; Jansen, M.; Kern, K.; Amsharov, K. Synthesis of Wide Atomically Precise Graphene Nanoribbons from Para-Oligophenylene Based Molecular Precursor. *Carbon* **2014**, *77*, 1187–1190.

(6) Zhang, H.; Lin, H.; Sun, K.; Chen, L.; Zagranyarski, Y.; Aghdassi, N.; Duhm, S.; Li, Q.; Zhong, D.; Li, Y.; *et al.* On-Surface Synthesis of Rylene-Type Graphene Nanoribbons. *J. Am. Chem. Soc.* **2015**, *137*, 4022–4025.

(7) Talirz, L.; Söde, H.; Dumslaff, T.; Wang, S.; Sanchez-Valencia, J. R.; Liu, J.; Shinde, P.; Pignedoli, C. A.; Liang, L.; Meunier, V.; *et al.* On-Surface Synthesis and Characterization of 9-Atom Wide Armchair Graphene Nanoribbons. *ACS Nano* **2017**, *11*, 1380–1388.

(8) Kharche, N.; Meunier, V. Width and Crystal Orientation Dependent Band Gap Renormalization in Substrate-Supported Graphene Nanoribbons. *J. Phys. Chem. Lett.* **2016**, *7*, 1526–1533.

(9) Deniz, O.; Sánchez-Sánchez, C.; Dumslaff, T.; Feng, X.; Narita, A.; Müllen, K.; Kharche, N.; Meunier, V.; Fasel, R.; Ruffieux, P. Revealing the Electronic Structure of Silicon Intercalated Armchair Graphene Nanoribbons by Scanning Tunneling Spectroscopy. *Nano Lett.* **2017**, *17*, 2197–2203.

(10) Merino-Díez, N.; Garcia-Lekue, A.; Carbonell-Sanromà, E.; Li, J.; Corso, M.; Colazzo, L.; Sedona, F.; Sánchez-Portal, D.; Pascual, J.I.; de Oteyza, D. G. Width-Dependent Band Gap in Armchair Graphene Nanoribbons Reveals Fermi Level Pinning on Au(111). *ACS Nano* **2017**, 11, 11661–11668.

(11) Ruffieux, P.; Wang, S.; Yang, B.; Sánchez-Sánchez, C.; Liu, J.; Dienel, T.; Talirz, L.; Shinde, P.; Pignedoli, C. A.; Passerone, D.; *et al.* On-Surface Synthesis of Graphene Nanoribbons with Zigzag Edge Topology. *Nature* **2016**, *531*, 489–492.

(12) Wakabayashi, K.; Sasaki, K.; Nakanishi, T.; Enoki, T. Electronic States of Graphene Nanoribbons and Analytical Solutions. *Sci. Technol. Adv. Mater.* **2010**, *11*, 054504.

(13) Wang, S.; Talirz, L.; Pignedoli, C. A.; Feng, X.; Müllen, K.; Fasel, R.; Ruffieux, P. Giant Edge State Splitting at Atomically Precise Graphene Zigzag Edges. *Nat. Commun.* **2016**, *7*, 11507.

(14) Tao, C.; Jiao, L.; Yazyev, O. V.; Chen, Y.-C.; Feng, J.; Zhang, X.; Capaz, R. B.; Tour, J. M.; Zettl, A.; Louie, S. G.; *et al.* Spatially Resolving Edge States of Chiral Graphene Nanoribbons. *Nat. Phys.* **2011**, *7*, 616–620.

(15) Pan, M.; Girão, E. C.; Jia, X.; Bhaviripudi, S.; Li, Q.; Kong, J.; Meunier, V.; Dresselhaus, M. S. Topographic and Spectroscopic Characterization of Electronic Edge States in CVD Grown Graphene Nanoribbons. *Nano Lett.* **2012**, *12*, 1928–1933.

(16) Han, P.; Akagi, K.; Federici Canova, F.; Mutoh, H.; Shiraki, S.; Iwaya, K.; Weiss, P. S.; Asao, N.; Hitosugi, T. Bottom-Up Graphene-Nanoribbon Fabrication Reveals Chiral Edges and Enantioselectivity. *ACS Nano* **2014**, *8*, 9181–9187.

(17) Sánchez-Sánchez, C.; Dienel, T.; Deniz, O.; Ruffieux, P.; Berger, R.; Feng, X.; Müllen, K.; Fasel, R. Purely Armchair or Partially Chiral: Noncontact Atomic Force Microscopy Characterization of Dibromo-Bianthryl-Based Graphene Nanoribbons Grown on Cu(111). *ACS Nano* **2016**, *10*, 8006–8011.

(18) Schulz, F.; Jacobse, P. H.; Canova, F. F.; van der Lit, J.; Gao, D. Z.; van den Hoogenband, A.; Han, P.; Klein Gebbink, R. J. M.; Moret, M.-E.; Joensuu, P. M.; *et al.* Precursor Geometry Determines the Growth Mechanism in Graphene Nanoribbons. *J. Phys. Chem. C* **2017**, *121*, 2896–2904.

(19) Han, P.; Akagi, K.; Federici Canova, F.; Shimizu, R.; Oguchi, H.; Shiraki, S.; Weiss, P. S.; Asao, N.; Hitosugi, T. Self-Assembly Strategy for Fabricating Connected Graphene Nanoribbons. *ACS Nano* **2015**, *9*, 12035–12044.

(20) de Oteyza, D. G.; García-Lekue, A.; Vilas-Varela, M.; Merino-Díez, N.; Carbonell-Sanromà, E.; Corso, M.; Vasseur, G.; Rogero, C.; Guitián, E.; Pascual, J. I.; *et al.* Substrate-Independent Growth of Atomically Precise Chiral Graphene Nanoribbons. *ACS Nano* **2016**, *10*, 9000–9008.

(21) Weiss, C.; Wagner, C.; Kleimann, C.; Rohlfing, M.; Tautz, F. S.; Temirov, R. Imaging Pauli Repulsion in Scanning Tunneling Microscopy. *Phys. Rev. Lett.* **2010**, *105*, 086103.

(22) Kichin, G.; Weiss, C.; Wagner, C.; Tautz, F. S.; Temirov, R. Single Molecule and Single Atom Sensors for Atomic Resolution Imaging of Chemically Complex Surfaces. *J. Am. Chem. Soc.* **2011**, *133*, 16847–16851.

(23) Kichin, G.; Wagner, C.; Tautz, F. S.; Temirov, R. Calibrating Atomic-Scale Force Sensors Installed at the Tip Apex of a Scanning Tunneling Microscope. *Phys. Rev. B* **2013**, *87*, 081408.

(24) Krejci, O.; Hapala, P.; Ondracek, M.; Jelinek, P. Principles and Simulations of High-Resolution STM Imaging with Flexible Tip Apex. *Phys. Rev. B* **2017**, *95*, 045407.

(25) Gross, L.; Moll, N.; Mohn, F.; Curioni, A.; Meyer, G.; Hanke, F.; Persson, M. High-Resolution Molecular Orbital Imaging Using a p-Wave STM Tip. *Phys. Rev. Lett.* **2011**, *107*, 086101.

(26) Söde, H.; Talirz, L.; Gröning, O.; Pignedoli, C. A.; Berger, R.; Feng, X.; Müllen, K.; Fasel, R.; Ruffieux, P. Electronic Band Dispersion of Graphene Nanoribbons via Fourier-Transformed Scanning Tunneling Spectroscopy. *Phys. Rev. B* **2015**, *91*, 045429

(27) Pascual, J. I.; Song, Z.; Jackiw, J. J.; Horn, K.; Rust, H.-P. Visualization of Surface Electronic Structure: Dispersion of Surface States of Ag(110). *Phys. Rev. B* **2001**, *63*, 241103(R)

(28) Ruffieux, P.; Cai, J.; Plumb, N. C.; Patthey, L.; Prezzi, D.; Ferretti, A.; Molinari, E.; Feng, X.; Müllen, K.; Pignedoli, C. A.; *et al.* Electronic Structure of Atomically Precise Graphene Nanoribbons. *ACS Nano* **2012**, *6*, 6930–6935.

(29) Linden, S.; Zhong, D.; Timmer, A.; Aghdassi, N.; Franke, J. H.; Zhang, H.; Feng, X.; Müllen, K.; Fuchs, H.; Chi, L.; *et al.* Electronic Structure of Spatially Aligned Graphene Nanoribbons on Au(788). *Phys. Rev. Lett.* **2012**, *108*, 216801.

(30) Nakada, K.; Fujita, M.; Dresselhaus, G.; Dresselhaus, M. S. Edge State in Graphene Ribbons: Nanometer Size Effect and Edge Shape Dependence. *Phys. Rev. B* **1996**, *54*, 17954–17961.

(31) Jiang, Z.; Song, Y. Band Gap Oscillation and Novel Transport Property in Ultrathin Chiral Graphene Nanoribbons. *Phys. B Condens. Matter* **2015**, *464*, 61–67.

(32) Shota, S.; Oshiyama, A. Energetics, Electron States, and Magnetization in Nearly Zigzag-Edged Graphene Nano-Ribbons. *J. Phys. Soc. Jap.* **2015**, *84*, 024704.

(33) Carvalho, A. R.; Warnes, J. H.; Lewenkopf, C. H. Edge Magnetization and Local Density of States in Chiral Graphene Nanoribbons. *Phys. Rev. B* **2014**, *89*, 245444.

(34) Raza, H.; Kan, E. C. Armchair Graphene Nanoribbons: Electronic Structure and Electric-Field Modulation. *Phys. Rev. B* **2008**, *77*, 245434.

(35) Carbonell-Sanromà, E.; Brandimarte, P.; Balog, R.; Corso, M.; Kawai, S.; Garcia-Lekue, A.; Saito, S.; Yamaguchi, S.; Meyer, E.; Sánchez-Portal, D.; *et al.* Quantum Dots Embedded in Graphene Nanoribbons by Chemical Substitution. *Nano Lett.* **2017**, *17*, 50–56.

(36) Horcas, I.; Fernández, R.; Gómez-Rodríguez, J. M.; Colchero, J.; Gómez-Herrero, J.; Baro, A. M. WSXM: A Software for Scanning Probe Microscopy and a Tool for Nanotechnology. *Rev. Sci. Instrum.* **2007**, *78*, 013705.

(37) Soler, J. M.; Artacho, E.; Gale, J. D.; García, A.; Junquera, J.; Ordejón, P.; Daniel Sánchez-Portal. The SIESTA Method for Ab Initio Order- N Materials Simulation. *J. Phys. Condens. Matter* **2002**, *14*, 2745.

(38) Klimeš, J.; Bowler, D. R.; Michaelides, A. Chemical Accuracy for the van Der Waals Density Functional. *J. Phys. Condens. Matter* **2010**, *22*, 022201.

# Supplementary Information:

# Unraveling the electronic structure of narrow atomically-precise chiral graphene nanoribbons

Néstor Merino-Díez [‡,†], Jingcheng Li [†,§], Aran Garcia-Lekue [‡,⊥], Guillaume Vasseur [‡,§], Manuel Vilas-Varela [∇], Eduard Carbonell-Sanromà [†], Martina Corso [†,§], J. Enrique Ortega [‡,§,*], Diego Peña [∇], Jose I. Pascual [†,⊥], and Dimas G. de Oteyza *,[‡,†,§,⊥].

[‡]Donostia International Physics Center (DIPC), 20018 San Sebastián-Donostia, Spain

[†]CIC nanoGUNE, Nanoscience Cooperative Research Center, 20018 San Sebastián-Donostia, Spain

[§]Centro de Física de Materiales (CSIC/UPV-EHU) - Materials Physics Center, 20018 San Sebastián-Donostia, Spain

[⊥]Ikerbasque, Basque Foundation for Science, 48013 Bilbao, Spain

[∇]Centro de Investigación Química Biolóxica e Materiais Meloculares (CIQUS) and Departamento de Química Orgánica, Universidad de Santiago de Compostela, 15782 Santiago de Compostela, Spain

[*]Departamento de Física Aplicada I, Universidad del País Vasco (UPV/EHU), 20018 San Sebastián-Donostia, Spain

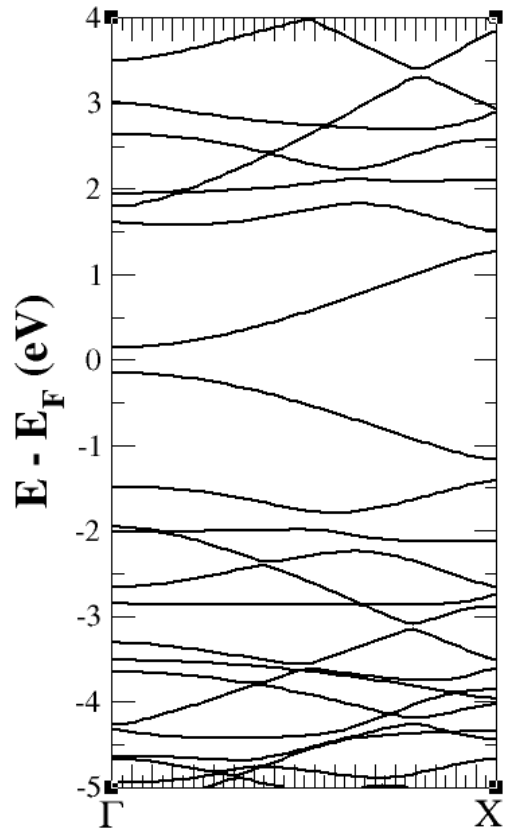

**Figure S1.** DFT calculated band structure for (3,1)-chGNRs on Au(111).

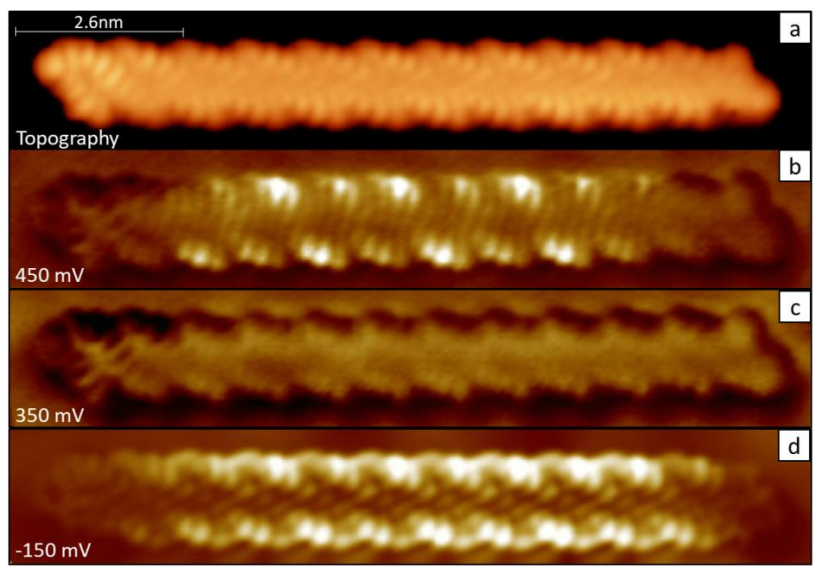

**Figure S2.** STM constant-current conductance maps evidencing the absence of signal at energies within the gap and the clear GNR-related signal at the VB and CB onsets. (a) STM topography image (13.1 nm x 3.1 nm; $V_s$ = -0.15V; $I_t$ = 1nA) and (b-d) STM constant-current conductance maps (13.1 nm x 3.1 nm; $I_t$ = 1nA) (b) at conduction band onset, (c) within the band gap and (d) at the valence band onset.

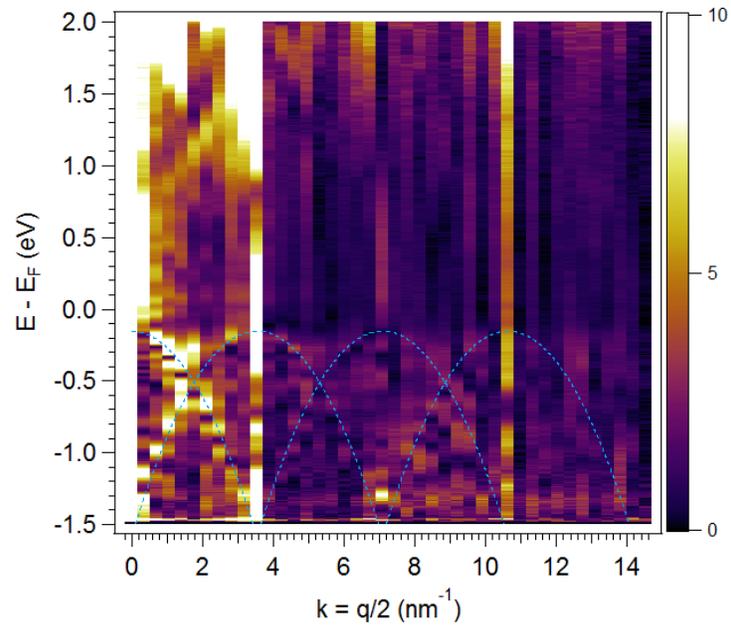

**Figure S3**. Line-by-line FT-STS data displayed in Figure 3i, depicting the complete energy and momentum range experimentally measured, and showing valence band replica along the momentum space.

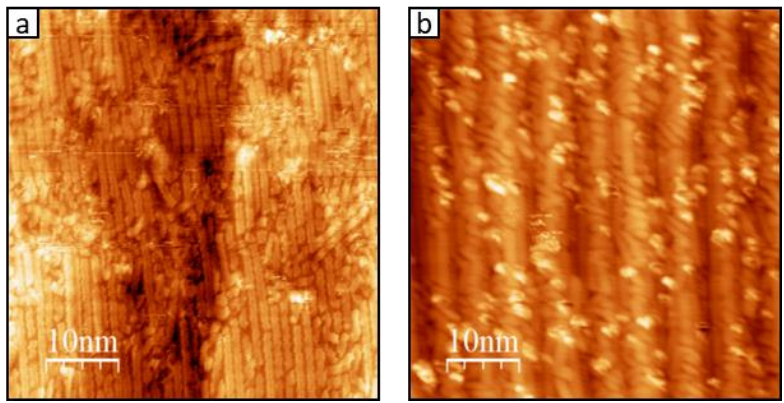

**Figure S4**. Constant current STM images of (3,1)-chGNRs grown on (a) Au(322) and (b) Au(788), evidencing the much more limited degree of order for the latter.

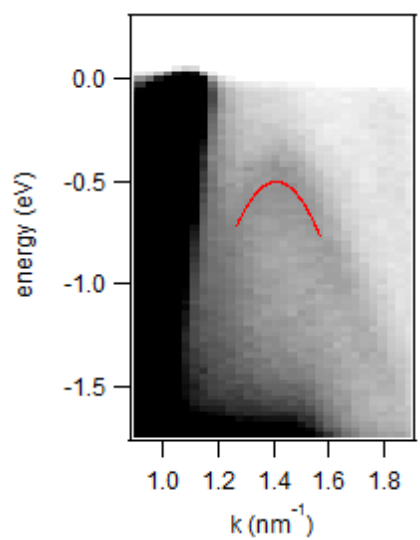

**Figure S5**. ARPES measurements displaying the raw photoemission intensity of the GNR´s valence band and the fit to the topmost band region for determination of the effective mass.

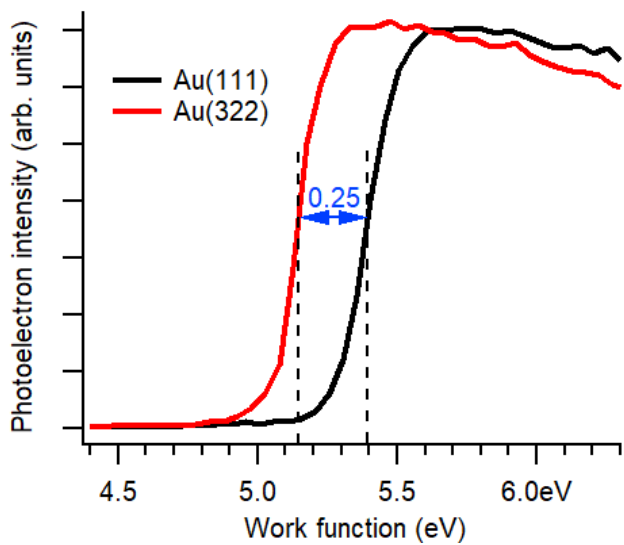

**Figure S6**. Work function measurements of Au(111) and Au(322) surfaces. Photoemission spectra were taken with the sample biased at -9 V and the absolute work function value was extracted from $\Phi = h\nu - (E_F - E_{cut\ off})$, where $E_F$ and $E_{cut\ off}$ have been taken from the corresponding maxima in derivated spectra. Spectra in the figure have been subsequently shifted to fit the cut off energy to the extracted work function values.